\documentclass{article}

\usepackage{amsthm,amsmath,natbib}
\usepackage[top=1in, bottom=1in, left=1in, right=1in]{geometry}

\usepackage{graphicx}
\usepackage{mathrsfs} 
\usepackage{tikz}
\usepackage{amssymb}
\usepackage{dsfont} 

\newcommand{\pd}[2]{\frac{\partial #1}{\partial #2}}

\newcommand{\od}[2]{\frac{\text{d} #1}{\text{d} #2}}

\newcommand{\dx}[1]{\ \text{d} #1}
\newcommand{\E}{\mathbb{E}}

\newcommand{\Y}{\mathbf{Y}}
\newcommand{\Z}{\mathbf{Z}}
\newcommand{\W}{\mathbf{W}}
\newcommand{\bbeta}{\boldsymbol{\beta}}
\newcommand{\bgamma}{\boldsymbol{\gamma}}

\newcommand{\indicator}[1]{\mathds{1}\{ #1 \}}

\title{Birth-death processes}
\author{Forrest W. Crawford\footnote{Department of Biostatistics, Yale School of Public Health}~ and Marc A. Suchard\footnote{Departments of Biomathematics, Human Genetics, and Biostatistics, University of California Los Angeles}}

\begin{document}

\maketitle









\begin{abstract}
\noindent Many important stochastic counting models can be written as general birth-death processes (BDPs).  BDPs are continuous-time Markov chains on the non-negative integers in which only jumps to adjacent states are allowed.  BDPs can be used to easily parameterize a rich variety of probability distributions on the non-negative integers, and straightforward conditions guarantee that these distributions are proper.  BDPs also provide a mechanistic interpretation -- birth and death of actual particles or organisms -- that has proven useful in evolution, ecology, physics, and chemistry.  Although the theoretical properties of general BDPs are well understood, traditionally statistical work on BDPs has been limited to the simple linear (Kendall) process, which arises in ecology and evolutionary applications.  Aside from a few simple cases, it remains impossible to find analytic expressions for the likelihood of a discretely-observed BDP, and computational difficulties have hindered development of tools for statistical inference.  But the gap between BDP theory and practical methods for estimation has narrowed in recent years.  There are now robust methods for evaluating likelihoods for realizations of BDPs: finite-time transition, first passage, equilibrium probabilities, and distributions of summary statistics that arise commonly in applications.  Recent work has also exploited the connection between continuously- and discretely-observed BDPs to derive EM algorithms for maximum likelihood estimation. Likelihood-based inference for previously intractable BDPs is much easier than previously thought and regression approaches analogous to Poisson regression are straightforward to derive. In this review, we outline the basic mathematical theory for BDPs and demonstrate new tools for statistical inference using data from BDPs.  We give six examples of BDPs and derive EM algorithms to fit their parameters by maximum likelihood.  We show how to compute the distribution of integral summary statistics and give an example application to the total cost of an epidemic. Finally, we suggest future directions for innovation in this important class of stochastic processes.
\end{abstract}




\section{Introduction}

\begin{figure}
\centering
\includegraphics{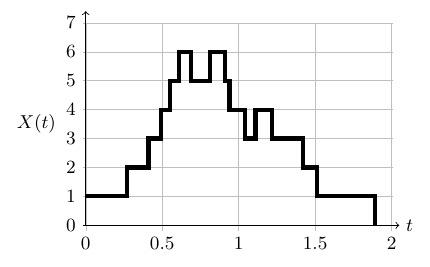}
\caption{Stochastic simulation of a BDP starting at $X(0)=1$ on the interval $0<t<2$.}
\label{fig:foo}
\label{fig:sim}
\end{figure}

Birth-death processes (BDPs) are a flexible class of continuous-time Markov chains that model the number of ``particles'' in a system, where each particle can ``give birth'' to another particle or ``die'' \citep{Feller1971Introduction,Karlin1975First}.  The rate of births and deaths at any given time depends on how many extant particles there are.  When there are $k$ particles, a birth occurs with instantaneous rate $\lambda_k$ and a death with instantaneous rate $\mu_k$.  In the classical ``simple linear'' BDP, $\lambda_k=k\lambda$ and $\mu_k=k\mu$ so that per-particle birth and death rates remain constant.  In a ``general'' BDP, $\lambda_k$ and $\mu_k$ can be any function of $k$ but are time-homogeneous \citep{Kendall1948Generalized,Kendall1949Stochastic}.  Table \ref{tab:examples} gives examples of well-known BDPs and their birth and death rates.  Figure \ref{fig:sim} shows an example realization from a BDP.

The usefulness of BDPs lies in the fact that ``particle'' can refer to a member of any discrete potentially interacting system in which one only keeps track of the number of objects in existence.  BDPs are popular modeling tools in evolution, population biology, genetics, and ecology \citep{Novozhilov2006Biological}.  For example, if we interpret the particles as species in a macro-evolutionary setting, BDPs can be used to study speciation and extinction over evolutionary timescales \citep{Nee1994Reconstructed,Nee2006Birth}.  BDPs can also be used to study infectious disease dynamics in a finite population, where the number of individuals infected is the quantity of interest \citep{Bailey1964Elements,Andersson2000Stochastic}. In molecular evolution, BDPs can model inserted and deleted nucleotides in a DNA or RNA sequence as part of a probabilistic alignment method \citep{Thorne1991Evolutionary,Holmes2001Evolutionary}, mobile/transposable genetic elements \citep{Rosenberg2003Estimating}, gene families \citep{Demuth2006Evolution}, or even whole chromosomes \citep{Mayrose2010Probabilistic}.  BDPs can model populations of organisms in a resource-limited environment \citep{Tan1991Stochastic,Renshaw1993Modelling,Renshaw2011Stochastic}. In finite populations, BDPs are commonly used to model quantities of interest in an evolutionary setting, such as allele frequencies, selection, or coalescence \citep{Moran1958Random,Krone1997Ancestral,Kingman1982Genealogy}.  

Many important models in queuing theory can be written as general BDPs \citep{Ross1995Stochastic,Norris1998Markov,Renshaw2011Stochastic}.  In basic Markovian queues, customers arrive into a queue or buffer as a Poisson process with rate $\lambda$, and waiting customers are served (removed from the queue) with per-customer service rate $\mu$. In the $M/M/\infty$ queue, also known as the immigration-death process, there are infinitely many servers, so the arrival and service (birth and death) rates are $\lambda_k=\lambda$ and $\mu_k = k\mu$ for $k>0$.  In the $M/M/1$ queue, also known as the immigration-emigration model, there is only a single server, so the rates are $\lambda_k =\lambda$ and $\mu_k =\mu$.  In the $M/M/c$ queue, there are exactly $c$ servers, so $\mu_k=\min\{c,k\}\mu$. 

BDPs can also be useful for defining arbitrary probability distributions on the non-negative integers.  \citet{Crawford2014Counting} demonstrate that any sum of exchangeable Bernoulli random variables can be exactly represented as a pure-birth BDP.  In fact, \citet{Faddy1997Extended} shows that one can define a pure birth process (a BDP with death rates $\mu_k=0$ for all $k$) whose transition probabilities reproduce any discrete distribution on the counting numbers.  \citet{Klar2010Zipf} establish a correspondence between several power law distributions and the long-time limit of specially constructed BDPs, providing a time-dependent interpretation that may be useful for modelling mechanistic processes that give rise to power law outcomes.  \citet{Crawford2012Transition} define a BDP to mimic a process of frameshift-aware insertions and deletions in DNA sequences.  \citet{Lee2011Using} set the birth and death rates of a BDP to exhibit over-dispersion relative to the Poisson distribution, and \citet{Crawford2014Sex} define a BDP to model rounding in counts of sex partners to multiples of 5, 10, 25, or 50 in self-reported counts of sex partners in a public health study.

There is a rich history of theoretical research into the properties of BDPs.  \citet{Kendall1948Generalized,Kendall1949Stochastic} introduce the process with constant per-particle birth and death rates and finds the transition probabilities by a generating function argument.  In their groundbreaking series of papers, \citeauthor{Karlin1957Differential} analyze properties of BDPs, including stationary distributions, moments, transition probabilities, recurrence and passage times, and other quantities of interest \citep{Karlin1957Differential,Karlin1957Classification}.  They also explore in-depth applications of this theory to BDPs whose rates depend linearly on $k$ \citep{Karlin1958Linear}, and queuing processes \citep{Karlin1958Many}.  

Beyond the work of \citeauthor{Karlin1957Differential}, many authors have discovered extensions and deeper interpretations for the theoretical properties of BDPs.  For example, the theory of BDPs is intimately related to properties of continued fractions \citep{Guillemin1999Excursions}.  \citet{Flajolet2000Formal} elucidate the relationship between sample trajectories (or state paths) of a BDP and lattice path combinatorics via continued fractions and develop expressions for a variety of recurrence and passage time variables in terms of continued fractions.  \citet{Lenin2000Birth} and \citet{Parthasarathy1998Birth} discuss further some well-known continued fractions whose connection to BDPs previously went unappreciated.

\begin{table}
  \centering
  \begin{tabular}{lcc}
    \hline
    Model                    & $\lambda_k$       & $\mu_k$ \\
    \hline
    Poisson                  & $\lambda$         & 0       \\
    Yule/Pure birth          & $k\lambda$        & 0       \\
    Survival/Pure death      & 0                 & $k\mu$  \\
    Kendall                  & $k\lambda$        & $k\mu$  \\
    Kendall $+$ immigration & $k\lambda+\alpha$ & $k\mu$   \\
    $M/M/1$ queue            & $\lambda$         & $\mu$   \\
    $M/M/c$ queue            & $\lambda$         & $\text{min}(k,c)\mu$  \\
    $M/M/\infty$ queue       & $\lambda$         & $k\mu$     \\
    SI/Logistic              & $k(N-k)\lambda$   & 0       \\
    SIS                      & $k(N-k)\lambda$   & $k\mu$       \\
    Moran/Ehrenfest          & $k(N-k)\lambda$   & $k(N-k)\mu$  \\
    \hline
  \end{tabular}
  \caption{Some well-known BDPs with birth and death rates $\lambda_k$ and $\mu_k$.  The SI and SIS models refers to the susceptible-infectious(-susceptible) process in epidemiology in which there are $k$ infectious individuals in a finite population of size $N$.  The Moran/Ehrenfest process models the change in the numbers of particles of two types, where transitions between types occur at a rate proportional to the number of potential contacts between members each type in a finite population of size $N$.}
  \label{tab:examples}
\end{table}

The study of BDPs has benefited from wide interest in the theoretical properties of this class of processes.  But their usefulness as flexible tools for statistical inference has been under-appreciated.  In this review, we outline basic properties of BDPs and show how to perform principled statistical inference using data from continuous and discrete observation of BDPs.  First, we present the basic time-evolution equations of general BDPs, derive the transition probabilities for the Kendall process \citep{Feller1971Introduction,Kendall1948Generalized}, and describe the analytic theory developed by \citet{Karlin1957Differential,Karlin1957Classification} for general BDPs.  Then we outline a computational strategy for evaluating BDP transition probabilities using a continued fraction representation of their Laplace transform, which allows routine computation of likelihoods for discretely observed processes \citep{Crawford2012Transition}.  We describe a generic class of EM algorithms for maximum likelihood (or maximum \emph{a posteriori}) inference for discretely observed BDPs \citep{Crawford2014Estimation}.  Finally, we derive the distribution of integral summary statistics of BDPs that arise often in applications.

\section{Background}

A BDP is a continuous-time Markov chain $X(t)$ counting the number of particles in a system at time $t$, taking values on the non-negative integers $\mathbb{N}$.  To construct a general BDP in a formal way, we must define the rules according to which the number of particles evolves. We do this by specifying the behavior of the process for a very short time $dt$, when there are $k$ particles in the system.  If $dt$ is very small, the probability of an event during $(t,t+dt)$ that occurs with rate $r$ is approximately $rdt$.  Therefore, the probability of a birth in the interval $(t,t+dt)$, given $X(t)=k$, is
\begin{equation}
  \Pr\big(X(t+dt)=k+1\mid X(t)=k\big) = \lambda_k dt + o(dt).
\end{equation}
Intuitively, this means that the probability of more than one birth event in a small time $d t$ is negligibly small.  The probability of a death in $(t,t+d t)$ is likewise
\begin{equation}
 \Pr\big(X(t+d t)=k-1\mid X(t)=k\big) = \mu_k d t + o(d t) , 
\end{equation}
where $k\geq 1$.
Together, these assumptions imply that the probability of no births or deaths occurring during $(t,t+d t)$ is
\begin{equation}
\Pr\big(X(t+dt)=k\mid X(t)=k\big) = 1 - (\lambda_k + \mu_k) d t + o(d t).
\end{equation}

\subsection{Transition probabilities}

Let $P_{ab}(t)=\Pr(X(t)=b\mid X(0)=a)$ be the transition probability from state $X(0)=a$ to $X(t)=b$.  We can use the above expressions to form a differential equation describing the change in transition probabilities over time. 
Suppose that $X(0)=a$.  At the current time $t$, we want to know the probability that in the next $d t$ units of time, the process will reach state $b$.  We look into the future by writing the probabilities of three types of events that can take the process to state $b$: birth from $b-1$, death from $b+1$, or no change from $b$:
\begin{equation}
  \begin{split}
P_{ab}(t+d t) &= \lambda_{b-1}P_{a,b-1}(t)d t + \mu_{b+1}P_{a,b+1}(t) d t \\
&\quad + (1 - \lambda_b - \mu_b)P_{ab}(t)d t + o(d t) .
\end{split}
\end{equation}
Subtracting $P_{ab}(t)$ from both sides, dividing by $d t$, and sending $d t$ to zero, we obtain the Kolmogorov forward equations:
\begin{equation}
\od{P_{ab}(t)}{t} = \lambda_{b-1}P_{a,b-1}(t) + \mu_{b+1}P_{a,b+1}(t) - (\lambda_b + \mu_b)P_{ab}(t) ,
\label{eq:odesintro}
\end{equation}
where $P_{ab}(0)=1$ if $a=b$ and zero otherwise.  In this article, we always assume $\mu_0=\lambda_{-1}=0$; this keeps the process on the non-negative integers.  Letting $\mathbf{P}(t) = \{ P_{ab}(t) \}$ 
in matrix form, \eqref{eq:odesintro} becomes
\begin{equation}
  \od{\mathbf{P}(t)}{t} = \mathbf{A} \mathbf{P}(t),
\label{eq:matrixode}
\end{equation}
where $\mathbf{A}$ is the infinitesimal generator matrix with entries $\mathbf{A}=\{a_{ij}\}$, $a_{i,i-1}=\mu_i$, $a_{ii}=-(\lambda_i+\mu_i)$, and $a_{n,n+1}=\lambda_i$.  In the matrix case, the initial condition becomes $\mathbf{P}(0)=\mathbf{I}$.   This infinite sequence of coupled ordinary differential equations 
can be difficult or impossible to solve for many general BDPs \citep{Novozhilov2006Biological,Renshaw2011Stochastic}.  

\subsubsection{Kendall process}

In the simple linear BDP, also known as the Kendall process where $\lambda_k=k\lambda$ and $\mu_k=k\mu$, it is possible to solve for these transition probabilities explicitly by finding a generating function solution to the forward equations \citep{Bailey1964Elements,Lange2010Applied}.  To illustrate, let $G_a(s,t) = \sum_{k=0}^\infty s^k P_{ak}(t)$.  Let $b=k$ in \eqref{eq:odesintro}, multiply both sides by $s^k$, and sum on $k$ to obtain
\begin{equation}
\begin{split}
 \pd{G_a(s,t)}{t} &= \sum_{k=0}^\infty s^k \od{P_{ak}(t)}{t} \\
  &= \lambda s^2 \sum_{k=1}^\infty (k-1)s^{k-2} P_{a,k-1}(t)
     + \mu \sum_{k=0}^\infty (k+1)s^k P_{a,k+1}(t) \\
  &\qquad   - (\lambda+\mu) s\sum_{k=0}^\infty ks^{k-1} P_{ak}(t) \\
  &= (\lambda s - \mu)(s-1) \pd{G_a(s,t)}{s} ,
\end{split}
\label{eq:Ga}
\end{equation}
with the initial condition $G_a(s,0) = s^a$.  
The solution is 
\begin{equation}
 G_a(s,t) = \left(\frac{\mu(s-1) + (\lambda s - \mu) e^{-(\lambda-\mu)t}}{\lambda(s-1) + (\lambda s - \mu) e^{-(\lambda-\mu)t}} \right)^a. 
\end{equation}
Inverting and finding the $b$th coefficient of the power series $G_a(s,t)$, we find the transition probabilities  
\begin{equation}
 P_{ab}(t) = \sum_{j=0}^{\min(a,b)} \binom{a}{j} \binom{a+b-j-1}{a-1}\alpha^{a-j}\beta^{b-j}(1-\alpha-\beta)^j,
 \label{eq:kendallprob}
\end{equation}
where 
\begin{equation}
\alpha(t) = \frac{\mu(e^{(\lambda-\mu)t)} -1)}{\lambda(e^{(\lambda-\mu)t} -\mu)} \quad\text{and}\quad \beta(t) = \frac{\lambda(e^{(\lambda-\mu)t)} -1)}{\lambda e^{(\lambda-\mu)t} -\mu}.
\end{equation}

\subsubsection{General BDPs}

The problem becomes much more complicated for general BDPs.  \citet{Karlin1957Differential} present the definitive treatment of the existence of transition probabilities and other properties of BDPs.  They obtain the following integral form for the transition probabilities:
\begin{equation}
  P_{ab}(t) = \omega_b \int_0^\infty e^{-xt} Q_a(x) Q_b(x) \dx{\psi(x)},
  \label{eq:integralintro}
\end{equation}
where $\omega_0 = 1$ and $\omega_k = (\lambda_0\cdots\lambda_{k-1})/(\mu_1\cdots\mu_k)$ for $k \geq 1$.  Here, $Q_k(x)$, $k=0,1,2,\ldots$ is a system of orthogonal polynomials and $\psi(x)$ is an orthogonalizing spectral measure that are specific to a particular set of birth and death rates.  

This integral representation 
is intuitively satisfying because the time-dependency of $P_{ab}(t)$ is contained entirely in the exponential term, and $P_{ab}(t)$ depends on $Q_a(x)$ and $Q_b(x)$ in a simple way.  In addition, we have the obvious corollary that 
\begin{equation}
 \frac{P_{ab}(t)}{P_{ba}(t)} = \frac{\omega_b}{\omega_a} .
\end{equation}
Beyond these simple results related to the interpretation of \eqref{eq:integralintro}, the formalism developed by \citet{Karlin1957Differential} makes possible deep analytic insight into the behavior of general BDPs, including recurrence times and first passage times.  

\subsection{Equilibrium probabilities and explosion}

Equilibrium solutions are straightforward to obtain \citep{Renshaw2011Stochastic}.  Setting the left-hand side of the Kolmogorov forward equations \eqref{eq:odesintro} to zero and replacing the finite-time transition probabilities $P_{ab}(t)$ with the equilibrium probabilities $\pi_b$, we find that 
\begin{equation}
\mu_{b+1}\pi_{b+1} - \lambda_b\pi_b = \mu_b\pi_b - \lambda_{b-1}\pi_{b-1} .
\label{eq:equilibrecur}
\end{equation}
Since this is the case for every $b$, it is true for $b=0$ in particular, and $\mu_0=\lambda_{-1}=0$, so both sides of \eqref{eq:equilibrecur} are zero for every $b$ by induction.  This gives the detailed balance condition for continuous-time Markov chains,
\begin{equation}
 \mu_k \pi_k = \lambda_{k-1} \pi_{k-1}  \quad\text{for } k=1,2,\ldots .
\label{eq:detailedbalance}
\end{equation}
Therefore every general BDP is a reversible Markov chain.  
Iterating the recurrence \eqref{eq:detailedbalance}, we find that
\begin{equation}
\pi_k = \frac{\lambda_0\lambda_1\cdots\lambda_{k-1}}{\mu_1\mu_2\cdots\mu_k} \pi_0, 
\label{eq:equilib}
\end{equation}
where we have chosen $\pi_0$ so that $\sum_k \pi_k=1$.  Note that $\pi_k \propto \omega_k$ for every $k$.

The birth and death rates for a general BDP may be such that the process ``runs away'' to infinity in finite time.  This is 
known as explosive growth.  Formally, suppose the process begins at $X(0)=0$ and there are no absorbing states.  \citet{Renshaw2011Stochastic} shows that the expected first passage time to infinity $\tau^\infty$ is 
\begin{equation}
  \E(\tau^\infty\mid X(0)=0) = \sum_{j=0}^\infty \pi_j \sum_{n=j}^\infty \frac{1}{\mu_n \pi_n} ,
\label{eq:explode}
\end{equation}
where $\pi_1 = 1$ and 
\begin{equation} 
  \pi_n = \prod_{k=1}^n \frac{\lambda_{k-1}}{\mu_k} ,
\end{equation}
for $i>1$.  When \eqref{eq:explode} diverges, the process is non-explosive, and the first passage time from $0$ to any finite state $j$ is almost surely finite.  When \eqref{eq:explode} is finite, the first passage time to infinity is finite with non-zero probability.  

One result of special interest to us gives the conditions under which a BDP with a given generator $\mathbf{A}$ is unique: \citeauthor{Karlin1957Differential} show that there is only one transition probability matrix $\mathbf{P}(t)$ that satisfies \eqref{eq:matrixode} if and only if 
\begin{equation}
 \sum_{k=0}^\infty \left( \omega_k + \frac{1}{\lambda_k\omega_k} \right) = \infty .
\label{eq:honest}
\end{equation}
This property assures that probability is conserved on the non-negative integers.  We will always assume this is the case in what follows.

Despite the elegant representation \eqref{eq:integralintro} for the transition probabilities, it can be very difficult to find the polynomials $\{Q_k(x)\}$ \citep{Renshaw2011Stochastic,Novozhilov2006Biological}.  In addition, the problem of finding these polynomials and measure $\psi$ is a fundamentally analytical task, and is generally not amenable to computational solution.  In other words, one cannot simply compute $P_{ab}(t)$ using a computer for an arbitrary set of birth and death rates $\{\lambda_k\}$ and $\{\mu_k\}$ using the formula \eqref{eq:integralintro} alone.  For this reason, nearly all modeling applications use the simple linear BDP since it is analytically tractable.  \citet[][page 111]{Renshaw2011Stochastic} writes of the need for an alternative approach to solving the forward system in order to find transition probabilities for general BDPs:

\begin{quote}
``A worthwhile and potentially rewarding challenge would be to develop a simplified and user-friendly version of this technique which would work over a wide range of stochastic processes.''
\end{quote}
The next section is devoted to this task.


\section{Transition probabilities for general BDPs}

\label{sec:transprobs}

We now outline a method, first presented in \citet{Crawford2012Transition} and based on work by \citet{Murphy1975Some}, for numerically computing the transition probabilities for a general BDP with arbitrary birth and death rates.  To proceed, denote the Laplace transform of $P_{ab}(t)$ as
\begin{equation}
 f_{ab}(s) = \mathscr{L}\left[P_{ab}(t)\right](s) = \int_0^\infty e^{-st} P_{ab}(t) \dx{t} .
\end{equation}
Now, applying the Laplace transform to \eqref{eq:odesintro} with $a=0$, we have
\begin{equation}
\begin{split}
  s f_{00}(s) - P_{00}(0) &=  \mu_1 f_{01}(s) - \lambda_0 f_{00}(s) \text{, and} \\
  s f_{0b}(s) - P_{0b}(0) &=  \lambda_{n-1}f_{0,b-1}(s) + \mu_{b+1} f_{0,b+1}(s) - (\lambda_b + \mu_b)f_{0b}(s) 
 \end{split}
\label{eq:recur0}
\end{equation}
for $b\geq 1$.  Recalling that $P_{00}(0) = 1$ and $P_{0b}(0) = 0$ for $b\geq 1$, we rearrange \eqref{eq:recur0} to find
\begin{equation}
\begin{split}
  f_{00}(s) &= \frac{1}{s + \lambda_0 - \mu_1 \left(\frac{f_{01}(s)}{f_{00}(s)}\right) } \text{, and} \\
  \frac{f_{0b}(s)}{f_{0,b-1}(s)} &= \frac{\lambda_{b-1}}{s + \mu_b + \lambda_{b} - \mu_{b+1} \left(\frac{f_{0,b+1}(s)}{f_{0b}(s)}\right)}. 
 \end{split}
 \label{eq:recur2}
\end{equation}
By combining these recurrence relations, we obtain the generalized continued fraction 
\begin{equation}
  f_{00}(s) = \cfrac{1}{s+\lambda_0 - \cfrac{\lambda_0 \mu_1}{s+\lambda_1+\mu_1 - \cfrac{\lambda_1 \mu_2}{s+\lambda_2+\mu_2 - \cdots}}},
  \label{eq:cfrac1}
\end{equation}
that is an exact expression for the Laplace transform of the transition probability $P_{00}(t)$ \citep{Karlin1957Differential,Bordes1983Application,Guillemin1999Excursions,Flajolet2000Formal}.  Now define $a_1 = 1$, $a_n = -\lambda_{n-2}\mu_{n-1}$, $b_1 = s+\lambda_0$ and $b_n = s+\lambda_{n-1}+\mu_{n-1}$ for $n\geq 2$.  Then \eqref{eq:cfrac1} becomes
\begin{equation}
  f_{00}(s) = \frac{a_1}{b_1+} \frac{a_2}{b_2+} \frac{a_3}{b_3+} \cdots 
 \label{eq:cfrac3}
\end{equation}
in more concise notation.  We denote the $k$th convergent of the Laplace transform $f_{00}(s)$ by
\begin{equation}
 f_{00}^{(k)}(s) = \frac{a_1}{b_1+} \frac{a_2}{b_2+} \cdots \frac{a_k}{b_k} = \frac{A_k(s)}{B_k(s)}.
\label{eq:convergent}
\end{equation}
The main result of \citet{Crawford2012Transition} is the following theorem giving continued fraction expressions for the Laplace transform of the transition probability in a general birth-death process.
\newtheorem{thm:cf}{Theorem}
\begin{thm:cf}
\label{thm:cf}
The Laplace transform of the transition probability $P_{ab}(t)$ is given by 
\begin{equation}
f_{ab}(s) = \begin{cases} 
\displaystyle\left(\prod_{j=b+1}^a \mu_j\right)\frac{B_b(s)}{B_{a+1}(s)+} \frac{B_a(s) a_{a+2}}{b_{a+2}+} \frac{a_{a+3}}{b_{a+3}+}\cdots  & \text{for $b\leq a$}, \\
& \\
\displaystyle\left(\prod_{j=a}^{b-1} \lambda_j\right) \frac{B_a(s)}{B_{b+1}(s)+} \frac{B_b(s) a_{b+2}}{b_{b+2}+} \frac{a_{b+3}}{b_{b+3}+}\cdots  & \text{for $a \leq b$,}
\end{cases}
\label{eq:fmnfull}
\end{equation}
where $a_n$, $b_n$, and $B_n$ are as defined above.
\end{thm:cf}
The proof of this theorem relies on elementary manipulation of the continued fraction recurrences \eqref{eq:recur2}.  \citet{Crawford2012Transition} obtain time-domain transition probabilities $P_{ab}(t)$ from \eqref{eq:fmnfull} by numerically inverting the Laplace transforms.  We refer the reader to that publication for the computational details.  The method returns transition probabilities for many general BDPs that have eluded previous analytical and numerical methods.

\subsection{First passage times}

%
%

Now consider the time of first arrival of a BDP $X(t)$ into an arbitrary set $S$ of taboo states, and suppose $X(0)=i\in\mathbb{N}\setminus S$.  This first passage time is defined formally as
\begin{equation}
  \tau_i = \inf\ \{t: X(t) \in S \mid X(0)=i \} .
  \label{eq:fpt}
\end{equation} 
To find the relationship between first passage times and the expressions for transition probabilities discussed above, construct a new process $Y(t)$ identical to $X(t)$ except that $\lambda_j = \mu_j = 0$ for every $j\in S$, so every state in $S$ is absorbing.  Then for this modified process, with $P_{ij}(t) = \Pr(Y(t)=j\mid Y(0)=i)$,
\begin{equation}
  \Pr( \tau_i < t) = \sum_{j\in S} P_{ij}(t) .
\end{equation}
The intuitive reason for this equality is the absorbing nature of the states in $S$: if $Y$ reaches an absorbing state $j\in S$ at any time before $t$, we must have $Y(t)=j$.  Furthermore, $Y$ cannot visit more than one state in $S$, so the absorption events are mutually exclusive and the probability of absorption is simply the sum of the individual absorption probabilities.  Therefore the cumulative distribution function of the first passage time into $S$ is given by the sum of the transition probabilities from $i$ to every taboo state in $S$ for the modified process $Y(t)$.  


\section{Likelihoods}

One factor hindering more widespread adoption of BDPs by applied researchers is the difficulty in performing statistical estimation of the unknown parameters in a BDP using real-world data \citep{Holmes2001Evolutionary,Doss2013Fitting}.  Typically efforts in estimation for BDPs have been limited to continuous observation of the process \citep{Moran1951Estimation,Moran1953Estimation,Anscombe1953Sequential,Darwin1956Behaviour,Wolff1965Problems,Reynolds1973Estimating}.  In addition, most work to date has focused on the simple linear BDP because it is analytically tractable \citep{Keiding1975Maximum,Thorne1991Evolutionary,Dauxois2004Bayesian,Rosenberg2003Estimating}.  However, in practice researchers often observe data from BDPs only at discrete times through longitudinal sampling.  In addition, the simple linear BDP may be unappealing because it fails to capture more complicated dynamics of population growth and decay that arise when particles do not behave independently.  To learn from discretely-observed general BDPs, we will need more advanced statistical tools.

\subsection{Likelihood for the continuously-observed process}

In a discretely-observed general BDP, the likelihood cannot be written in closed form, making analytic maximum likelihood estimation impossible.  However, the likelihood of a continuously-observed BDP is straightforward to express \citep{Reynolds1973Estimating,Keiding1975Maximum}.  To develop the likelihood for continuously-observed data from a general BDP, we note the following important fact: the exponentially distributed waiting time of a continuous-time Markov process in a certain state is independent of the destination of the next jump \citep{Lange2010Applied}.  Recall that the waiting time $W$ for the first event to occur from state $k$ is exponentially distributed with rate $\lambda_k+\mu_k$.  If the waiting time in the current state $k$ is $W=\tau$, and the next change is a birth, 
\begin{equation}
\begin{split}
  \Pr(W=\tau,\text{birth} \mid X(0)=k) &= \Pr(W=\tau\mid X(0)=k)\Pr(\text{birth}\mid X(0)=k)  \\
   &= (\lambda_k+\mu_k) e^{-(\lambda_k+\mu_k)\tau}  \left(\frac{\lambda_k}{\lambda_k+\mu_k}\right) \\
   &= \lambda_k e^{-(\lambda_k+\mu_k)\tau}.
\end{split}
\end{equation}
Likewise, the probability of a waiting time $W=\tau$ followed by a death is 
\begin{equation}
 \Pr(W=\tau,\text{death} \mid X(0)=k) = \mu_k e^{-(\lambda_k+\mu_k)\tau}.
\end{equation}
Since we can only observe the process for a finite time $t$, the last observation will be the waiting time in some state $k$ from the time of the jump to $k$ to the end of observation.  Using the same reasoning, 
\begin{equation}
 \Pr(W\geq\tau \mid \text{no births or deaths},X(0)=k) = e^{-(\lambda_k+\mu_k)\tau}.
\end{equation}

To write the likelihood of a continuously-observed BDP from time $0$ to $t$, we introduce some notation to ease our presentation.  Suppose we observe $n$ jumps in the time interval $(0,t)$, and label the jumps $i=1,\ldots,n$.  Let $W_i$ be the waiting time in the current state just before the $i$th jump.  Define the indicator $B_i=1$ if the $i$th jump is a birth, and $B_i=0$ if the $i$th jump is a death.  Let $t_1,\ldots,t_n$ be the times of the $n$ jumps, with $t_0=0$ and $t_n<t$.  Then the likelihood of a sequence of observations $\Y=\{X(\tau),0<\tau<t\}$ is 
\begin{equation}
\begin{split}
L &=  \prod_{i=1}^n \Bigg[ \Pr\big(W_i=t_i-t_{i-1} \mid X(t_{i-1})\big)\\
  & \qquad \times \Pr\big(\text{birth}\mid X(t_{i-1}))^{B_i} \Pr(\text{death}\mid X(t_{i-1})\big)^{1-B_i} \Bigg] \\
  & \qquad \times \Pr\big(W_{n+1}=t-t_n\mid \text{no births or deaths},X(t_n)\big) \\\\
  &= \prod_{i=1}^n (\lambda_{X(t_{i-1})}+\mu_{X(t_{i-1})}) \exp\left[-(\lambda_{X(t_{i-1})}+\mu_{X(t_{i-1})})(t_i-t_{i-1})\right] \\
  &\qquad \times \left(\frac{\lambda_{X(t_{i-1})}^{B_i} \mu_{X(t_{i-1})}^{1-B_i} }{\lambda_{X(t_{i-1})}+\mu_{X(t_{i-1})}}\right) \times \exp\left[-(\lambda_{X(t_n)}+\mu_{X(t_n)})(t-t_n)\right] \\\\
  &= \prod_{i=1}^n \lambda_{X(t_{i-1})}^{B_i} \mu_{X(t_{i-1})}^{1-B_i} \exp\left[-(\lambda_{X(t_{i-1})}+\mu_{X(t_{i-1})})(t_i-t_{i-1})\right] \\
  &\qquad \times \exp\left[-(\lambda_{X(t_n)}+\mu_{X(t_n)})(t-t_n)\right] ,
\end{split}
\label{eq:likderiv}
\end{equation}
where $X(t_{i-1})$ is the state just before the $i$th jump. This cumbersome notation can be eliminated if we instead keep track of the total waiting time in each state and the number of births and deaths from each state.  Define $\indicator{E}$ to be the indicator of an event $E$, and let  
\begin{equation}
 T_k = \sum_{i=1}^n (t_i-t_{i-1})\indicator{X(t_{i-1})=k} 
\end{equation}
be the total time spent in state $k$ over all visits to $k$.  Then let
\begin{equation}
 U_k = \sum_{i=1}^n \indicator{X(t_{i-1})=k,B_i=1} 
\end{equation}
be the number of up steps (births) from state $k$, and let
\begin{equation}
 D_k = \sum_{i=1}^n \indicator{X(t_{i-1})=k,B_i=0}
\end{equation}
be the number of down steps (deaths) from state $k$.  Then we can re-write the likelihood \eqref{eq:likderiv} in much simpler and more transparent form as 
\begin{equation}
 L = \prod_{k=0}^\infty \lambda_k^{U_k} \mu_k^{D_k} \exp[-(\lambda_k+\mu_k)T_k] .
\label{eq:likcompact}
\end{equation}
Of course, in a BDP observed continuously for a finite time (for which \eqref{eq:honest} holds), there are only finitely many jumps observed, so the product above is not really infinite in practice.  

Equation \eqref{eq:likcompact} also reveals that the likelihood for a continuously-observed BDP is a member of the exponential family, where $\{U_k\}$, $\{D_k\}$, and $\{T_k\}$ for $k=0,1,\ldots$ are the sufficient statistics of the continuously-observed BDP likelihood.  In other words, one only needs to know the total number of up and down steps from, and time spent in, each state $k$ visited by the process in order to compute the likelihood.  


\subsection{Example: continuously-observed Kendall process}

Maximum likelihood estimation for continuously-observed BDPs is often straightforward.  Consider the simple linear BDP with birth rate $\lambda_k=k\lambda$ and death rate $\mu_k=k\mu$.  The likelihood \eqref{eq:likcompact} of a single observation becomes, up to a normalizing constant, becomes
\begin{equation}
 L \propto \lambda^{U} \mu^{D} \exp\left[-(\lambda+\mu) \int_0^t X(\tau) \dx{\tau}\right] ,
\label{eq:liksimple}
\end{equation}
where $U=\sum_k U_k$ is the total number of up steps (births), $D=\sum_k D_k$ is the total number of down steps (deaths) during the interval $(0,t)$, and 
\begin{equation}
 \int_0^t X(\tau)\dx{\tau} = \sum_{k=0}^\infty k T_k 
\end{equation}  
is the ``total particle time'' or total time lived by every particle that existed during the interval $(0,t)$.  Maximizing \eqref{eq:liksimple} with respect to the unknown parameters $\lambda$ and $\mu$, we obtain the maximum likelihood estimators 
\begin{equation}
\hat{\lambda} = \frac{U}{\displaystyle\int_0^t X(\tau) \dx{\tau}} \quad\text{and}\quad \hat{\mu} = \frac{D}{\displaystyle\int_0^t X(\tau) \dx{\tau}} ,
\label{eq:simplemle}
\end{equation}
first given by \citet{Reynolds1973Estimating}.  Although the estimators provided by \eqref{eq:simplemle} involve an integral over the state path of the process, the integrand is simply a step function that is fully observed over $(0,t)$.

\subsection{Likelihood for the discretely-observed process}

Suppose now that the process $X(\tau)$ is observed only discretely, once at time $0$ and again at time $t$, without loss of generality owing to the Markov assumption.  Let us label the state of the BDP at these times as $X(0)=a$ and $X(t)=b$.  Then given that $X(0)=a$, the probability that $X(t)=b$ is the transition probability $P_{ab}(t)$.  In section \ref{sec:transprobs} we outlined a method for numerically computing this probability for any general BDP.  If we regard the transition probability $P_{ab}(t)$ as a function of some unknown parameters $\theta$ which control the birth and death rates, writing $P_{ab}(t|\theta)$, then we have the \emph{likelihood} of our observation,
\begin{equation}
  L(\theta) = P_{ab}(t|\theta) .
  \label{eq:naivelik}
\end{equation}
In principle, we could numerically maximize the likelihood for discrete observations to find an estimate of $\theta$.  However, as the number of parameters increases, na\"{i}ve numerical optimization often suffers from poor convergence \citep{Doss2013Fitting}.  The difficulty in writing or computing the likelihood for discrete observations from BDPs has limited the usefulness of BDPs in applications.  

In contrast to the appealing analytic characterization \eqref{eq:likcompact} of the continuously-observed process likelihood, the discretely-observed process is hard to characterize.  To bridge this gap, it is helpful to view computation of the likelihood in the discretely-observed process as a missing data problem.  When a BDP is observed discretely, we do not know the sufficient statistics $\{U_k,D_k,T_k\}_{k=0}^\infty$.  This perspective suggests that we exploit analytic information about these statistics, conditional on the start and end states of the observed process.


\section{EM algorithms for maximum likelihood estimation}

In this section, we review the estimation machinery developed by \citet{Crawford2014Estimation} for maximum likelihood or maximum \emph{a posteriori} estimation in BDPs.  When a BDP is discretely sampled, $U_k$, $D_k$, and $T_k$ are unobserved for every $k$; we cannot maximize the likelihood without knowing these statistics.  We therefore appeal to the expectation-maximization (EM) algorithm for iterative maximum likelihood estimation with missing data \citep{Dempster1977Maximum}.  When the incomplete data likelihood is intractable but the complete data likelihood has a simple form, the EM algorithm operates by replacing each missing datum by a conditional expectation as follows.  If $X$ is the complete (unobserved data), $Y$ represents the incomplete (observed) data, and $\ell(\theta|X)$ is the complete data log-likelihood, we form a surrogate function $Q$ as the expectation of the complete data likelihood, conditional on the observed data $Y$ and the current ($m$th) parameter iterate:
\begin{equation}
 Q\left(\theta\mid\theta^{(m)}\right) = \E\left(\ell(\theta|X)\mid Y=y, \theta^{(m)}\right). 
\end{equation}
This is the E-step of the EM algorithm, and it accomplishes a minorization of $\ell(\theta|X)$ at $\theta^{(m)}$.  The M-step maximizes (or takes a step toward the maximum of) $Q$.  By alternating these steps --- minorizing $\ell$ by $Q$, then finding a $\theta$ that increases $Q$ --- the EM algorithm drives succeeding iterates toward the MLE.  

Taking the expectation of the logarithm of \eqref{eq:likcompact}, conditional on the observed data $Y=(X(0)=a,X(t)=b,t)$ and the current parameter estimate $\theta^{(m)}$, we write the surrogate function for the BDP as follows:
\begin{equation}
\begin{split}
Q\big(\theta\mid\theta^{(m)}\big) &= \E\big[\ell(\theta) \mid Y, \theta^{(m)}\big] \\
 &= \sum_{k=0}^\infty \E(U_k|Y)\log\big[\lambda_k(\theta)\big] + \E(D_k|Y)\log\big[\mu_k(\theta)\big] \\
 &\qquad - \E(T_k|Y) \big[\lambda_k(\theta) + \mu_k(\theta)\big], 
\end{split}
\label{eq:Q}
\end{equation}
In the above equation and many that follow, we omit the dependence of the conditional expectations on $\theta^{(m)}$ from the $m$th iterate for visual clarity.  

To calculate the conditional expectations necessary for the E-step of the EM algorithm, we appeal to the following integral expressions
\begin{subequations}
\label{eq:integralexpectations}
	\begin{align}
\E(U_k|Y) &= \frac{\displaystyle\int_0^t P_{ak}(\tau)\lambda_k P_{k+1,b}(t-\tau)\dx{\tau}}{P_{ab}(t)}, \label{eq:euk} \\
 \E(D_k|Y) &= \frac{\displaystyle\int_0^t P_{ak}(\tau)\mu_k P_{k-1,b}(t-\tau)\dx{\tau}}{P_{ab}(t)}, \quad\text{and} \label{eq:edk} \\
 \E(T_k|Y) &= \frac{\displaystyle\int_0^t P_{ak}(\tau) P_{kb}(t-\tau)\dx{\tau}}{P_{ab}(t)} . \label{eq:etk} 
\end{align}
\end{subequations}
These expressions have appeared repeatedly in literature on inference for discretely-observed continuous-time Markov chains \citep{Lange1995Gradient,Holmes2002Expectation,Bladt2005Statistical,Hobolth2005Statistical,Metzner2007Generator}.  When the process takes only finitely many states, matrix solutions are possible using the uniformization method \citep{Neuts1995Algorithmic}.  \citet{Hobolth2009Simulation} develop efficient Monte Carlo methods using simulation conditioned on the start and end points of the discrete observation $Y$.  Finally, \citet{Doss2013Fitting} study a linear BDP on an infinite state space and derive the expectations analytically using a generating function argument.  None of the exact methods is a general approach for arbitrary BDPs on an infinite state space.  The Monte Carlo approaches, while not reliant on a particular parameterization of the process, can suffer from poor performance when observed realizations occur with low probability.  The lack of a reliable method for computing the E-step of the EM algorithm for discretely-observed BDPs has hindered progress on statistical inference for these processes.

An alternative approach introduced by \citet{Crawford2014Estimation} takes advantage of the Laplace transforms $f_{ab}(s)$ of the transition probabilities \eqref{eq:fmnfull}.  The numerators in \eqref{eq:integralexpectations} are time-domain convolutions of transition probabilities.  The functional form of these expressions suggests using the Laplace convolution property to obtain
\begin{subequations}
\label{eq:convolutionexpectations}
\begin{align}
\E(U_k|Y) &= \lambda_k \frac{\mathscr{L}^{-1}\Big[f_{ak}(s)\ f_{k+1,b}(s) \Big](t)}{P_{ab}(t)}, \label{eq:eukconv} \\
\E(D_k|Y) &= \mu_k \frac{\mathscr{L}^{-1}\Big[f_{ak}(s)\ f_{k-1,b}(s) \Big](t)}{P_{ab}(t)},\quad\text{and} \label{eq:edkconv}\\
\E(T_k|Y) &= \frac{\mathscr{L}^{-1}\Big[f_{ak}(s)\ f_{kb}(s) \Big](t)}{P_{ab}(t)},  \label{eq:etkconv}
\end{align}
\end{subequations}
where $\mathscr{L}^{-1}[\cdot]$ denotes inverse Laplace transformation.  These expressions are formally equivalent to \eqref{eq:integralexpectations}, but they offer substantial computational time savings over numerical integration of \eqref{eq:integralexpectations}, and make possible efficient computation of conditional expectations for EM algorithms for any BDP \citep{Crawford2014Estimation}. 

We now show how to complete the M-step for several BDP models.  The first two, variations on the simple linear (Kendall) process, were given in \citet{Crawford2014Estimation}.  The others are novel, yet remarkably easy to derive and implement computationally.  In each case, we describe the surrogate likelihood function $Q\big(\theta|\theta^{(m)}\big)$ and give the M-step updates for each unknown parameter.

\subsection{Example: discretely-observed Kendall process}

\label{sec:simple}

In the simple linear BDP, 
births and deaths happen at constant per-particle rates, so $\lambda_k = k\lambda$ and $\mu_k = k\mu$. The unknown is $\theta = (\lambda, \mu)$.  The surrogate function $Q$ becomes
\begin{equation}
\begin{split}
  Q(\theta) &= \sum_{k=0}^\infty \E(U_k|Y)\log[k\lambda] + \E(D_k|Y)\log[k\mu] - \E(T_k|Y) k(\lambda+\mu).
\end{split}
  \label{eq:qsimple}
\end{equation}
Maximizing \eqref{eq:qsimple} with respect to the $\theta$ yields the updates:
\begin{subequations}
\label{eq:simpleupdates}
	\begin{align}
		\lambda^{(m+1)} &= \frac{\E(U|Y)}{\E(T_\text{particle}|Y) }\text{ and}  \label{eq:simplelambda} \\
    \mu^{(m+1)} &= \frac{\E(D|Y)}{\E(T_\text{particle}|Y) },  \label{eq:simplemu} 
	\end{align}
\end{subequations}
where 
\begin{equation}
 T_\text{particle} = \int_0^t X(\tau) \dx{\tau} ,
\end{equation}
and we have again suppressed the dependence of the conditional expectations on $\theta^{(m)}$ for typographic clarity.  These expressions are identical in form to the estimators given in \eqref{eq:simplemle}, but are instead iterative updates in the EM algorithm.

\subsection{Example: linear BDP with immigration}

\label{sec:im}

The linear BDP with immigration is similar to the simple linear BDP, but there is a source of new arrivals whose rate is constant and does not depend on the number of particles already in existence.  This yields the birth and death rates $\lambda_k=k\lambda+\nu$ and $\mu_k=k\mu$.  The log-likelihood becomes
\begin{equation}
   \ell(\theta) = \sum_{k=0}^\infty U_k\log(k\lambda + \nu) + D_k\log(\mu) - T_k[k(\lambda +\mu) + \nu ].
	\label{eq:qim}
\end{equation}
Unfortunately, it is difficult to maximize the resulting surrogate function analytically.  But since each term in the sum is a concave function of the unknown parameters, we can separate them in a second minorizing function $H$ such that for all $\theta$, $H\big(\theta|\theta^{(m)}\big) \leq \ell(\theta)$ and $H\big(\theta^{(m)}|\theta^{(m)}\big) = \ell\big(\theta^{(m)}\big)$.  To accomplish the minorization, note that 
\begin{equation}
  \begin{split}
 \log(k\lambda+\nu) &\geq \frac{k\lambda^{(m)}}{k\lambda^{(m)}+\nu^{(m)}} \log\left[\frac{k\lambda^{(m)}}{k\lambda^{(m)}+\nu^{(m)}} \lambda \right] \\
 &\quad + \frac{\nu^{(m)}}{k\lambda^{(m)}+\nu^{(m)}} \log\left[ \frac{\nu^{(m)}}{k\lambda^{(m)}+\nu^{(m)}} \nu\right].
 \end{split}
\end{equation}
 We form a minorizing log-likelihood function $H$ as follows: 
\begin{equation}
\begin{split}
H\big(\theta|\theta^{(m)}\big) &= \sum_{k=0}^\infty U_k \big[ p_k \log\big(p_k \lambda\big) + (1-p_k)\log\big( (1-p_k) \nu \big)\big] + D_k\log(\mu) \\
        &\qquad- \big[k(\lambda+\mu) + \nu \big]T_k, 
\end{split}
  \label{eq:loglikim}
\end{equation}
where
\begin{equation}
p_k = \frac{k\lambda^{(m)}}{k\lambda^{(m)} + \nu^{(m)}} .
\label{eq:pk}
\end{equation}
Exploiting this
surrogate function and maximizing with respect to the unknown sufficient statistics gives the updates
\begin{subequations}
\label{eq:imupdates}
\begin{align}
\lambda^{(m+1)} &= \frac{\displaystyle\sum_{k=0}^\infty p_k \E(U_k|Y)}{\E(T_\text{particle}|Y)}\text{, and} \label{eq:imlambda} \\
\nu^{(m+1)} &= \frac{\displaystyle\sum_{k=0}^\infty (1-p_k) \E(U_k|Y)}{t}.  \label{eq:imnu}
\end{align}
\end{subequations}
The update for $\mu$ is the same as \eqref{eq:simplemu}.

\subsection{Example: pure-birth and generalized Poisson processes}

Recall that the Poisson process with arrival rate $\lambda$ is a BDP with $\lambda_k=\lambda$, $\mu_k=0$ for all $k$.  Many researchers have found that real-world count data are sometimes over- or under-dispersed relative to the Poisson distribution.  Statisticians seeking a more flexible distribution for count outcomes that can accommodate over- and under-dispersion have arrived at several alternative distributions.  A notable example that fits neatly into the BDP framework is the general pure-birth process with arbitrary birth rates $\lambda_k$, $k=0,1,\ldots$, and $\mu_k=0$ for all $k$.  This class of processes has an appealing property: it can recover any discrete probability distribution on the counting numbers by appropriately setting the birth rates \citep{Faddy1997Extended,Faddy2001Likelihood}.  \citet{Crawford2014Counting} show that any such pure-birth process can be represented as a sum of exchangeable Bernoulli random variables, a result that connects BDPs with phenomenological models often used for dependent outcomes in toxicology and epidemiology.  \citet[][page 65]{Renshaw2011Stochastic} gives an analytic form for these transition probabilities
\begin{equation}
P_{ab}(t) = \left(\prod_{k=a}^{b-1}\lambda_k\right) \sum_{k=a}^b \left(\prod_{\ell\neq k}(\lambda_\ell-\lambda_k)\right)^{-1} \exp[-\lambda_k t]
\label{eq:purebirth}
\end{equation}
for $0\le a \le b$ and $t>0$ provided that $\lambda_i \neq \lambda_j$ for all $i$ and $j$.  
While \eqref{eq:purebirth} has an appealing form, it depends on none of the birth rates being equal.  Another potentially serious drawback is that it can be numerically troublesome to compute; the summands may be alternating in sign and the product of small differences in the denominators can lead to serious roundoff error.  In many scenarios, especially when some observed counts are large and some $\lambda_k$'s are nearly or exactly equal, \eqref{eq:purebirth} provides an unappealing way to compute the likelihood.
Exactly equal $\lambda_k$ may arise, for example, when entertaining a Bayesian non-parametric prior.
Fortunately, the EM framework does not require use of \eqref{eq:purebirth}.  We now provide an example of a pure birth process intended to generalize the Poisson distribution to accommodate over- and under-dispersion. 

\citet{Faddy1997Extended} describes a class of pure-birth BDPs with $\lambda_k=\lambda(\gamma+k)^c$ and $\mu_k=0$, where $c=0$ corresponds to a Poisson process with rate $\lambda$, $c>0$ results in overdispersion relative to Poisson, and $c<0$ results in underdispersion. The log-likelihood for the continuously-observed process beginning at $X(0)=a$ and ending at $X(t)=b$ is 
\begin{equation}
\ell(\theta) = \sum_{k=a}^b \indicator{k<b}[\log(\lambda) + c\log(\gamma+k)] - \lambda(\gamma+k)^c T_k .
\end{equation}
Letting $\theta = (\lambda,\gamma,c)$, the surrogate function is 
\begin{equation}
  Q(\theta) = \sum_{k=a}^b \indicator{k<b}[\log(\lambda) + c\log(\gamma+k)] - \lambda(\gamma+k)^c \E(T_k|\Y). 
 \end{equation}
The update for $\lambda$ is given by
\begin{equation}
\lambda^{(m+1)} = \frac{b-a-1}{\displaystyle\sum_{k=a}^b (\gamma+k)^c \E(T_k|\Y)} ,
\end{equation}
but the updates for $\gamma$ and $c$ are not available in closed form. However, \citet{Lange1995Gradient} shows that one step of a gradient ascent algorithm suffices to preserve the ascent property of the EM algorithm.  Therefore a Newton-Raphson update can be derived, and 
\begin{equation}
  \begin{pmatrix} \gamma^{(m+1)} \\ c^{(m+1)} \end{pmatrix} = 
  \mathbf{d}^2 Q(\gamma^{(m)},c^{(m)})^{-1} \ \nabla Q(\gamma^{(m)},c^{(m)}) ,
\end{equation}
where $\nabla Q$ and $\mathbf{d}^2Q$ are the gradient and Hessian of $Q$ with respect to $\gamma$ and $c$ respectively.


\subsection{Example: Moran model}

The Moran process models genetic drift in a finite population by keeping track of the number of alleles of a certain type at a biallelic locus in a haploid population of constant size $N<\infty$.  Call the two alleles $A$ and $B$, and suppose we wish to keep track of the number of $A$ carriers in the population.  In the Moran model with selection, carriers of $A$ have fitness $\alpha$, and carriers of $B$ have fitness $\beta$.  For the sake of identifiability in a statistical setting, we specify $\beta=1$ and let $\alpha$ denote the relative fitness of $A$ carriers over $B$ carriers.  Furthermore, $A$ mutates to $B$ in one generation with probability $u$, and \emph{vice versa} with probability $v$.  When an existing individual dies, a new allele is drawn at random.  The birth and death rates are
\begin{equation}
 \begin{split}
    \lambda_n &= \frac{N-n}{N}\left[\alpha\frac{n}{N}(1-u) + \frac{N-n}{N} v \right] \text{, and} \\
        \mu_n &= \frac{n}{N}\left[\frac{N-n}{N}(1-v) + \alpha \frac{n}{N} u\right]
\end{split}
\label{eq:moranbdrates}
\end{equation}
for $n=0,\ldots,N$.  Forming the surrogate function from \eqref{eq:Q}, we see that maximizing the log-likelihood with respect to the unknowns $\alpha$, $u$, and $v$ is difficult.  However, we can again construct a minorizing function to separate the parameters in the logarithm terms.  We minorize the birth rate as
\begin{equation}
\begin{split}
 \log(\lambda_n) &\propto \log\left[n\alpha(1-u) +  (N-n) v \right] \\
    &\geq p_n^{(m)}\log\left(p_n^{(m)}n\alpha(1-u)\right) + (1-p_n^{(m)})\log\left((1-p_n^{(m)})(N-n)v\right) \\
 & \propto p_n^{(m)}\big(\log(\alpha)+ \log(1-u)\big) + (1-p_n^{(m)})\log(v) ,
\end{split}
\label{eq:moranminorize}
\end{equation}
where 
\begin{equation} 
p_n^{(m)} = \frac{n\alpha^{(m)}(1-u^{(m)})}{n\alpha^{(m)}(1-u^{(m)})+(N-n)v^{(m)}} .
\label{eq:moranpn}
\end{equation}
Although \eqref{eq:moranminorize} and \eqref{eq:moranpn} may appear complicated, this minorization has the effect of separating the parameters $\alpha$ and $u$ in the surrogate function, allowing closed-form updates.  In a similar way, we minorize the death rate as
\begin{equation}
  \log(\mu_n) 
 \propto q_n^{(m)} \log(1-v) + (1-q_n^{(m)})\big(\log(\alpha) + \log(u) \big) ,
\end{equation}
where 
\begin{equation} 
q_n^{(m)} = \frac{(N-n)(1-v^{(m)})}{(N-n)(1-v^{(m)})+n\alpha^{(m)}u^{(m)}} .
\end{equation}
We form the complete minorizing function $H$ as
\begin{equation}
\begin{split}
H(\theta) &= \sum_{k=0}^N B_k\Big[ p_k^{(m)}\big(\log(\alpha) + \log(1-u)\big) + (1-p_k^{(m)})\log(v)\Big] \\
   &\qquad + D_k\Big[ q_k^{(m)}\log(1-v) + (1-q_k^{(m)})\big(\log(\alpha) + \log(u)\big) \Big] \\
   &\qquad - \frac{T_k}{N^2}\Big[(N-k)k\alpha (1-u) + (N-k)^2 v + (N-k)k(1-v) + k^2\alpha u \Big] ,
\end{split}
\end{equation}
and the surrogate function is $Q(\theta) = \E(H(\theta)|\Y,\theta^{(m)})$.  A simple way to proceed is to find updates for each of the unknowns individually, conditional on the previous ($m$th) estimate of the others, giving a cyclic coordinate ascent algorithm.  The update for $\alpha$ is
\begin{equation}
\alpha^{(m+1)} = \frac{\displaystyle\sum_{k=0}^N p_k^{(m)} B_k + (1-q_k^{(m)}) D_k}{\displaystyle\frac{1}{N^2}\sum_{k=0}^N T_k\big[(N-k)k(1-u^{(m)}) + k^2u^{(m)}\big] }.
\end{equation}
The update for $u$ is the positive solution of the quadratic equation
\begin{equation}
  \begin{split}
 0 &= \sum_{k=0}^N -uB_k p_k^{(m)} + (1-u)D_k (1-q_k^{(m)}) \\
   &\quad  - u(1-u) \frac{T_k}{N^2}\big[ k^2 \alpha^{(m)} - (N-k)k\alpha^{(m)} \big] ,
\end{split}
\end{equation}
when $0<u<1$. The update for $v$ is obtained by similar manipulations.


\subsection{Example: maximum \emph{a posteriori} estimation for the Kendall process}

In a Bayesian setting, a prior distribution $f(\theta)$ on the unknown parameters $\theta$ is given, and we seek to maximize the log-posterior distribution of the parameters, given the data, $\Pr(\theta\mid Y) \propto \Pr(Y\mid \theta) f(\theta)$ to obtain the maximum \emph{a posteriori} (MAP) estimate of $\theta$.  Here the surrogate function becomes $Q\big(\theta|\theta^{(m)}\big) = \E\left(\ell(\theta)|Y,\theta^{(m)}\right) + \log\left[f\left(\theta\right)\right]$.  To illustrate, suppose that independent observations from a BDP follow the simple linear model, and we believe  that $\lambda$ and $\mu$ are \emph{a priori} independent and are Gamma-distributed:
\begin{equation}
 \lambda \sim \text{Gamma}(k_\lambda, \beta_\lambda) \qquad\text{and}\qquad \mu \sim \text{Gamma}(k_\mu, \beta_\mu) .
\end{equation} 
Then the unknowns are $\theta=(\lambda,\mu)$ and the log-prior for $\theta$ is
\begin{equation}
  \log f(\theta) \propto (k_\lambda-1)\log(\lambda) + (k_\mu-1)\log(\mu) 
  - \beta_\lambda \lambda - \beta_\mu \mu
\end{equation}
Ignoring irrelevant terms, the surrogate function becomes
\begin{equation}
  \begin{split}	
  Q\big(\theta|\theta^{(m)}\big) &= \E(U|Y)\log(\lambda) + \E(D|Y)\log(\mu) -  \E(T_\text{particle}|Y)(\lambda + \mu) \\
 &\quad + (k_\lambda-1)\log(\lambda) + (k_\mu-1)\log(\mu) 
 - \beta_\lambda \lambda - \beta_\mu \mu
\end{split}
\label{eq:qmap}
\end{equation}
The MAP updates are
\begin{subequations}
\label{eq:mapupdates}
\begin{align}
  \lambda^{(m+1)} &= \frac{\E(U|Y) + k_\lambda - 1}{\E(T_\text{particle}|Y) + 
  \beta_\lambda},
  \quad\text{ and} \label{eq:maplambda} \\
  \mu^{(m+1)} &= \frac{\E(D|Y) + k_\mu - 1}{\E(T_\text{particle}|Y) + 
  \beta_\mu}.
  \label{eq:mapmu}
  \end{align}
\end{subequations}

\subsection{Example: regression for count data}

Perhaps the most interesting use of EM algorithms for BDP inference is to provide a unified framework for regression estimation.  To illustrate, consider a collection of $n$ independent BDPs, $X^i(t)$ with $\lambda^i_k = \exp[\Z_i\beta]$ and $\mu_k^i=0$ for $i=1,\ldots,n$, where $\Z_i$ is a $d\times 1$ vector of covariates and $\bbeta$ is a covariate vector of corresponding dimension and $\mu_k=0$ for all $k$.  Then letting $X^i(0)=0$ and $X^i(1)=x_i$ for each $i$, the log-likelihood becomes
\begin{equation}
    \ell(\bbeta) = \sum_{i=1}^n x_i \Z_i\bbeta - \exp[\Z_i\bbeta] .
\end{equation}
This is the log-likelihood for classical Poisson regression, and updates are found using a Newton-Raphson step \citep{Dobson2001Introduction}.

It is possible to formulate an analogous model for the Kendall process.  Let $\lambda^i_k = k\exp[\Z_i\bbeta]$ and let $\mu^i_k=k\exp[\W_i\bgamma]$ be the birth and death rates of a BDP $X^i(t)$.  The log-likelihood is
\begin{equation}
  \ell(\bbeta,\bgamma) \propto \sum_{i=1}^n \sum_{k=0}^\infty U^i_k \Z_i\bbeta + D^i_k\W_i\bgamma - T^i_k\left( \exp[\Z_i\bbeta] + \exp[\W_i\bgamma]\right) ,
\end{equation}
where the statistics $U_k^i$, $D_k^i$, and $T_k^i$ correspond to observation $i$. When the process is discretely-observed, we form the surrogate as in Section \ref{sec:simple}, and find the gradient vector
\begin{equation}
    \nabla_\beta Q = \sum_{i=1}^n \E[U^i|Y_i]\Z_i - \E[T^i_\text{particle}|Y_i] \Z_i \exp[\Z_i\bbeta]  
\end{equation}
for $\bbeta$. The Hessian matrix is 
\begin{equation}
    d^2_\beta Q = -\sum_{i=1}^n \E[T^i_\text{particle}|Y_i] \Z_i \Z_i' \exp[\Z_i\bbeta]  .
\end{equation}
Then, the Newton-Raphson update for $\bbeta$ becomes
\begin{equation}
    \bbeta^{(m)} = \bbeta^{(m-1)} - \left( d_\beta^2 Q\right)^{-1} \nabla_\beta Q .
    \label{eq:kendallregressionupdate}
\end{equation}
A similar update is available for $\bgamma$.  We contrast the simplicity of the update expressions \eqref{eq:kendallregressionupdate} with the formula for the Kendall process transition probability \eqref{eq:kendallprob}.


\section{Integral functionals of BDPs}

\begin{figure*}
    \centering
    \includegraphics[width=0.9\textwidth]{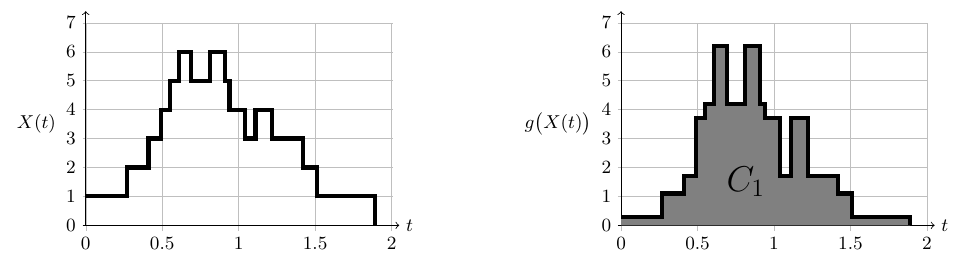}
    \caption{Illustration of the integral of a functional of a general birth-death process (BDP).  On the left, a BDP begins at $X(0)=1$ and ends when the process reaches the absorbing state $0$ just before time $t=2$.  On the right, $C_1 = \int_0^{\tau_1} g\big(X(t)\big)\dx{t}$ is the area under the trajectory of $g(X(t))$, where $g:\mathbb{N}\to [0,\infty)$ is an arbitrary positive ``reward'' or ``cost'' function.  The upper limit of integration $\tau_1$ is the first passage time to zero, beginning at $X(0)=1$.  }
\label{fig:integral}
\end{figure*}

Many important real-life applications of BDPs can be characterized as questions about the distribution of summary statistics.  A common feature of stochastic processes in decision-making contexts is that the parameters estimated by the statistical inference procedure are not always the ones of interest in the application.   Often the quantity of interest is a summary statistic related to the time-integral of the process.  To illustrate, let $g:\mathbb{N}\to [0,\infty)$ be a function and let $S$ be a set of ``taboo'' or prohibited states.  Suppose the initial state of the BDP is $X(0)=i \in \mathbb{N}\setminus S$.  Define the functional
\begin{equation}
  C_i = \int_0^{\tau_i} g\big(X(t)\big) \dx{t} ,
  \label{eq:Ci}
\end{equation}
where the upper limit of integration is the first passage time
\begin{equation}
  \tau_i = \inf\ \{t: X(t) \in S \mid X(0)=i \}. 
\end{equation}
Here, $C_i$ is a functional because it maps a realization of the stochastic process $g\big(X(t)\big)$ to its integral.  Figure \ref{fig:integral} shows an example realization of a BDP and its integral $C_i$ with $S=\{0\}$.  The left-hand side shows a BDP beginning at $X(0)=1$, and ending at $X(\tau_1)=0$.  The right-hand plot shows $g\big(X(t)\big)$ over the same time interval, and the area under the trajectory is $C_i$.

Expressions like \eqref{eq:Ci} arise often in applied work.  For example, epidemiologists usually estimate the parameters (contact/infection rate and recovery rate) of an epidemic process from data, but their objective is to make inference of the predictive distribution of the \emph{cost} of the epidemic in the future.  Operations researchers may estimate the arrival rate $\lambda$ and service rate $\mu$ in a queuing process, but the object of inference is the distribution of \emph{customer-hours} waited.  Traffic engineers may be interested in the number of \emph{vehicle-hours} waited in models for highway accident delays \citep{Gaver1969Highway}.  

To illustrate the role of integral summaries of BDPs in statistical prediction, let $p(c|\theta)$ be the density of $C_i$ given $\theta$.  The posterior predictive uncertainty about the statistic is the marginal distribution 
\begin{equation}
  \begin{split}
  p(c|Y) &= \int_\Theta p(c|\theta) p(\theta|Y) \dx{\theta} \\
         &\propto \int_\Theta p(c|\theta) p(Y|\theta) p(\theta) \dx{\theta} , \\
\end{split}
\label{eq:marginal}
\end{equation}
where $p(\theta|Y)$ is the sampling distribution of $\theta$ given the realized data $Y$.  In a Bayesian context, $p(\theta|Y)$ is a posterior distribution, and we might estimate $p(c | Y)$ by a Monte Carlo approximation involving $N$ draws $\theta_j \sim p(\theta|Y)$ via
\begin{equation}
  p(c|Y) \approx \frac{1}{N} \sum_{j=1}^N p(c|\theta_j). 
\end{equation}

\subsection{Background on integrals of BDPs}

\citet{Karlin1957Classification,Karlin1957Differential} provided the first theoretical tools for working with integral functionals of general BDPs.  \citet{Puri1966Homogeneous,Puri1968Some} derives the characteristic function for the joint distribution of simple linear BDP and its integral and gives expressions for moments and limiting distributions \citep{Puri1971MethodI,Puri1972MethodII,Puri1972MethodIII}.  \citet{McNeil1970Integral} gives the first results for general BDPs, \citet{Gani1971Joint} derive expressions for the joint distribution of a general BDP and its integral, and \citet{Kaplan1974Limit} provides limit theorems for integrals of simple BDPs with immigration.  Straightforward methods for moments of integrals of general BDPs using Laplace transforms  are also available \citep{Hernandez1999Basic,Pollett2003Method,Pollett2003Integrals,Gani2008Simple}.  However, most analyses of integral functionals of general BDPs are limited to simple analytically tractable models or focused on moments.



Now we consider the problem of computing the distribution of \eqref{eq:Ci}.  Our emphasis on first-passage times as the upper limit of integration in \eqref{eq:Ci} has two benefits.  First, our analyses need not be conditional on an arbitrary time in the future.  Second, first passage times allow us to exploit powerful analytic tools that establish a correspondence between transition probabilities and first-passage times, enabling analytic progress on integrals for arbitrary well-behaved processes.  Our presentation follows the outline given by \citet{McNeil1970Integral}.  Let $c_i(s) = \E\left[e^{-sC_i}\right]$ be the Laplace transform of $C_i$.  Note that if $X(0)=i\in S$ then $\tau_i=0$, $C_i = 0$, and so $c_i(s)=1$.  Now by an analogous conditioning argument for $X(0)=i \notin S$, we re-write the Laplace transform as
\begin{equation}
  \begin{split}
    c_i(s) &= \int_0^\infty \E\left[e^{-s(C_{i+1} + ug(i))}\right] \Pr(\text{birth},U=u\mid X(0)=i) \dx{u} \\
                  &\quad + \int_0^\infty \E\left[e^{-s(C_{i-1} + ug(i))}\right]\Pr(\text{death},U=u\mid X(0)=i)  \dx{u} \\
                  &= \E\left[e^{-s C_{i+1}}\right] \int_0^\infty e^{-s ug(i)} \lambda_i e^{-(\lambda_i + \mu_i)u} \dx{u} \\
                  &\quad + \E\left[e^{-s C_{i-1}}\right] \int_0^\infty e^{-s ug(i)} \mu_i e^{-(\lambda_i + \mu_i)u} \dx{u} \\
                  &= c_{i+1}(s) \lambda_i \int_0^\infty e^{-u(s g(i) + \lambda_i + \mu_i)} \dx{u} \\
                  &\quad +  c_{i-1}(s) \mu_i \int_0^\infty e^{-u(s g(i) + \lambda_i + \mu_i)} \dx{u} ,
\end{split}
\end{equation}
that gives
\begin{equation}
  \big(s g(i) + \lambda_i + \mu_i\big) c_i(s) =  \lambda_i c_{i+1}(s) + \mu_i c_{i-1}(s) .
\end{equation}
Now dividing both sides of the above by $g(i)$, we find that 
\begin{equation}
 \big(s + \lambda_i^* + \mu_i^*\big) c_i(s) = \lambda_i^* c_{i+1}(s) + \mu_i^* c_{i-1}(s) ,
 \label{eq:bwd}
\end{equation}
where $\lambda_i^* = \lambda_i/g(i)$ and $\mu_i^* = \mu_i/g(i)$.  Therefore, we see that \eqref{eq:bwd} is simply the backward equation for a modified process with birth and death rates $\lambda_i^*$ and $\mu_i^*$ for $i\in\mathbb{N}$. The forward equation for the cumulative distribution function of $c_i$ is therefore equivalent to \eqref{eq:odesintro} with the modified birth and death rates.

\citet{Pollett2003Integrals} gives the conditions, analogous to those for \eqref{eq:explode}, under which this modified process explodes.  We note that differentiation of solutions of \eqref{eq:bwd} yields the moments of $C_i$, as noted by \citet{McNeil1970Integral} and subsequently refined by \citet{Hernandez1999Basic}, \citet{Stefanov2000Note}, and \citet{Pollett2003Integrals}.  We refer interested readers to those papers and focus here on results for the distribution of $C_i$ that are more useful in statistical and decision applications.  

To take advantage of \eqref{eq:bwd}, we modify \eqref{eq:fpt} as follows.  Fix $S\subset \mathbb{N}$ and suppose $X(t)$ is a general BDP with rates $\{\lambda_n\}$ and $\{\mu_n\}$ with starting state $X(0)=i \in \mathbb{N}\setminus S$.  Suppose $g(n)$ is a positive function defined for all $n\in\mathbb{N}$.  Let $Y(t)$ be a general BDP with rates $\lambda^*_n=\lambda_n/g(n)$ and $\mu^*_n=\mu_n/g(n)$ for all $n\in \mathbb{N}\setminus S$, and $\lambda^*_n=\mu_n^*=0$ for every $n\in S$.  Then let $P_{ij}^*(t) = \Pr(Y(t)=j\mid Y(0)=i)$.  We then have 
\begin{equation}
  H(c) = \Pr( C_i < c) = \sum_{j\in S} P^*_{ij}(c). 
  \label{eq:integralcdf}
\end{equation}
If instead of the cumulative distribution function $H(c)$ of $C_i$, we wish to have the probability density, we could numerically differentiate \eqref{eq:integralcdf}.  However, using the properties of the Laplace transform, 
\begin{equation}
  \begin{split}
    h(c) &= \od{}{c} \Pr( C_i < c) \\
         &= \sum_{j\in S} \od{}{c} P^*_{ij}(c) \\ 
         &= \sum_{j\in S} \mathscr{L}^{-1}\Big[ s f^*_{ij}(s) - P^*_{ij}(0) \Big](c) \\
         &= \sum_{j\in S} \mathscr{L}^{-1}\Big[ s f^*_{ij}(s) \Big](c) 
\end{split}
  \label{eq:integraldensity}
\end{equation}
where $f^*_{ij}(s)$ is the Laplace transform of $P^*_{ij}(t)$, $\mathscr{L}^{-1}[\cdot]$ denotes Laplace inversion, and $P^*_{ij}(0)=0$ for all $j\in S$ since we have assumed $i\notin S$.

\subsection{Example: probabilistic control of an epidemic} 

\label{sec:sis}

In infectious disease epidemiology, stochastic modeling can give valuable insight into both disease dynamics and optimal intervention strategies \citep{Wickwire1977Mathematical,Ball1986Unified}.  The total cost of an infectious disease epidemic is proportional to the area under the time trajectory of the number of infected people \citep{Jerwood1970Note,Gani1972Cost}.  To illustrate, we model the number of infected persons in a homogeneously mixing population as a type of general BDP.  This simple model, called the susceptible-infected-susceptible (SIS) model, keeps track of the number of infected in a finite population of size $N$ \citep{Bailey1957Mathematical}.  If there are currently $n<N$ infected persons in the population, the rate of new infections is proportional to the product of the number infected $n$ and susceptible $N-n$.  The contact/transmission rate between infected and susceptible persons is $\lambda$.  Infected persons recover and revert to susceptible status with constant per-person rate $\mu$.  For a SIS process $X(t)$, the addition and removal rates are 
\begin{equation}
\lambda_n =\lambda n (N-n) \qquad\text{and}\qquad \mu_n = n (\mu+\epsilon) ,
\label{eq:sisrates}
\end{equation}
where $\epsilon$ is a positive control parameter related to vaccination or some other public health intervention strategy.  Suppose the initial number of infected is $X(0)=i \leq N$ and we are interested in the total cost of the epidemic until its eventual extinction, so $S=\{0\}$.  Let the cost of managing the epidemic per unit time be $a\epsilon$.  Additionally, let the cost per infected person per unit time be $b>0$, so the cost function becomes $g(n)=a\epsilon+bn$.  Then the total cost is
\begin{equation}
  C_i = \int_0^{\tau_i} \big[ a\epsilon + bX(t) \big] \dx{t} = a\epsilon\tau_i + b\int_0^{\tau_i} X(t)\dx{t} ,
\end{equation}
where $\tau_i$ is the time to extinction of the epidemic.  

\begin{figure}
  \begin{center}
    \includegraphics[width=0.6\textwidth]{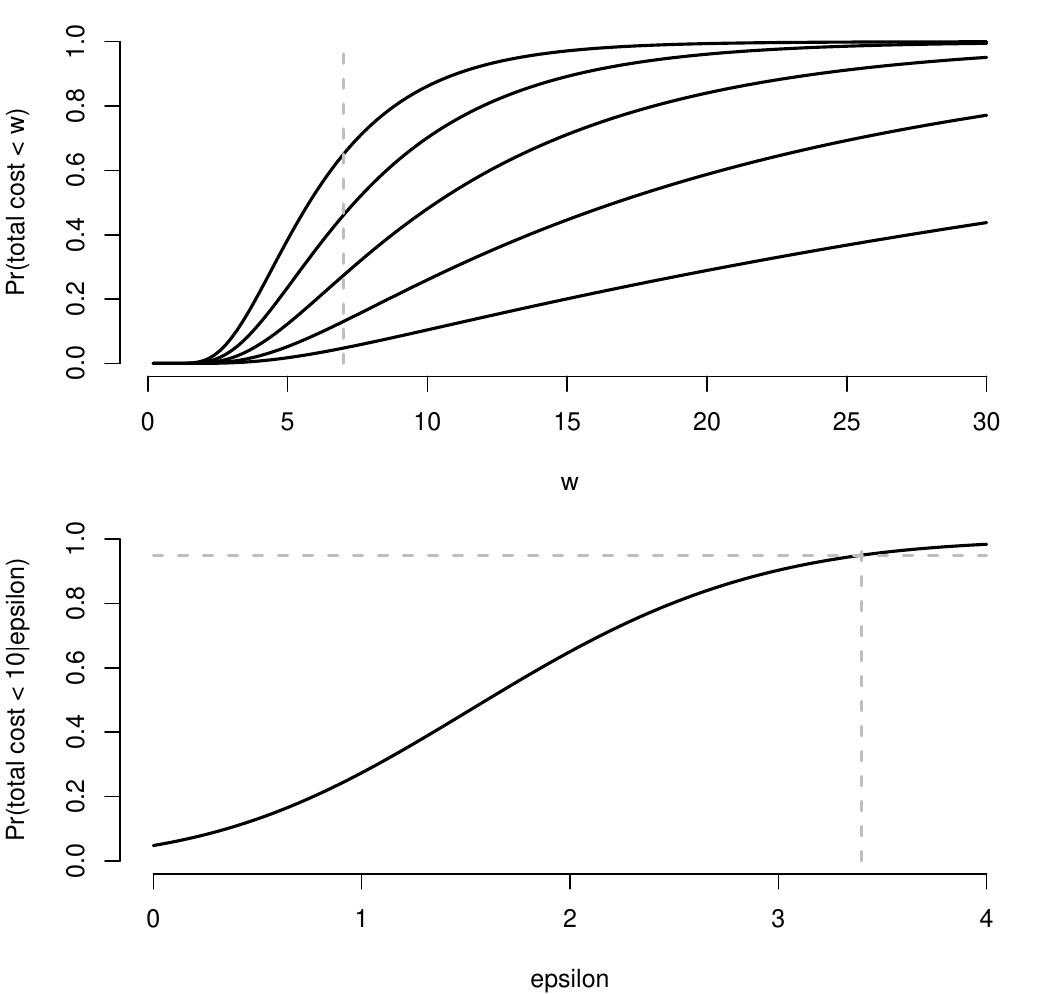}
  \end{center}
  \caption{Probabilistic control of a stochastic SIS epidemic.  At top, the distribution of total epidemic cost $C_i$ for different values of a control parameter $\epsilon$.  The dashed gray vertical line is at $w=7$, and we wish to keep $C_i<7$ with high probability.  At bottom, the probability that $C_i<7$ as a function of the control parameter $\epsilon$.  The horizontal gray dashed line denotes 0.95, and the vertical dashed line is the smallest epsilon that achieves $\Pr(C_i<7) > 0.95$; this yields $\epsilon \approx 3.4$.  In this way, we can easily find the smallest value of a control parameter that bounds the probability that the epidemic will exceed a certain threshold. }
  \label{fig:sis}
\end{figure}

Most optimal control models seek a policy that minimizes the expected total cost, corresponding to the expectation of \eqref{eq:Ci} under certain conditions on the intervention and cost functions \citep{Lefevre1981Optimal,Cai1994Stochastic,Clancy1999Optimal,Guo2009Continuous}.  
The availability of probability distributions for the total cost allows us to seek the minimal intervention policy that guarantees that the total cost of the epidemic is small with high probability.  Let $X(t)$ be the process with rates given by \eqref{eq:sisrates} for a certain control setting $\epsilon$.  Then we wish to find the smallest $\epsilon$ such that
\begin{equation}
  \Pr\left(C_i < c \right) < 1-\alpha  ,
  \label{eq:controlcost}
\end{equation}
where $c$ is a desired bound on the total cost, and $0<\alpha<1$ is a small probability.  
Assuming this probability is continuous and increases monotonically with $\epsilon$ near $1-\alpha$, it is straightforward to find the smallest $\epsilon$ that satisfies \eqref{eq:controlcost}.  

Figure \ref{fig:sis} shows how to find the minimal $\epsilon$ for a SIS process with $N=100$ individuals, $X(0)=50$, infectivity $\lambda=0.1$, recovery rate $\mu=8$, control cost $a=0.1$, and per-infected cost $b=0.3$ per unit time.  The top traces show the cumulative distribution function of the total cost for $\epsilon=0,0.5,1,1.5,2$.  The vertical gray line shows $C_i=7$, and we wish to keep the total cost less than 7 with probability $1-\alpha=0.95$.  The bottom trace shows $\Pr(C_i<7)$ as a function of $\epsilon$.  The horizontal gray dashed line shows 0.95 probability, and the vertical gray dashed line shows the smallest value of $\epsilon$ ($\epsilon\approx 3.4$) that achieves this bound.  



\section{Discussion: likelihood-based inference for BDPs}

BDPs are vital tools for modeling stochastic counting processes in epidemiology, evolution, ecology, chemistry, physics, and other fields.  Modeling with BDPs is often straightforward; by considering rates of addition of new particles and removal of existing particles, conditional on the number already present, researchers can specify the birth and death rates $\{\lambda_k,\mu_k\}_{k=0}^\infty$.  The ease of modeling with BDPs stands in stark contrast to the difficulty of inference using stochastic realizations of BDPs.  Routine use of BDPs in statistical settings has been thwarted by intractable likelihoods and burdensome computations.  A unified perspective on BDPs with arbitrary birth and death rates has remained elusive, until recently. 

Laplace transforms of transition probabilities provide the essential analytic tools for bridging this gap in practice.  Our approach for computing transition probabilities (likelihoods) in \eqref{eq:fmnfull} and conditional expectations in the E-step \eqref{eq:convolutionexpectations} is general, robust, and computationally efficient.  Laplace transforms of first-passage times also play an important role in finding the distribution of integral functionals of BDPs in applications.  As a theoretic tool, this Laplace-perspective is not new; \citet{Karlin1957Classification,Karlin1957Differential,Karlin1958Linear} discuss the fundamental importance of Laplace transforms for analysis of BDPs.  More recent results related to combinatorial properties of BDPs also rely on Laplace transforms \citep{Guillemin1998Continued,Guillemin1999Excursions} and \citet{Flajolet2000Formal}.

In this article, we have outlined new tools for practical likelihood-based analysis inference of BDP parameters under discrete and continuous observation of the process.  In particular, BDP generalizations of Poisson regression yield more flexible and easy-to-fit models for count data.  We have intentionally limited our discussion to basic computation of likelihoods, algorithms for maximum likelihood estimation, and finding the distribution of integral summary statistics for general BDPs.  But these are only the first steps toward a comprehensive theory of estimation for BDPs.  Ideally, we would like to see an analysis of identifiability, consistency and other statistical properties, like the groundbreaking work of \citet{Guttorp1991Statistical} for Galton-Watson branching processes.  We hope this review will stimulate statistical research related to BDPs with a view to bringing this rich class of stochastic models into wider use by applied scientists.






\section*{Acknowledgements}

FWC was supported by NIH grant T32GM008185 and startup funds from the Yale School of Public Health.
MAS was supported by NIH grants R01 AI107034 and R01 HG006139, and NSF grant DMS 1264153 and IIS 1251151.


\bibliographystyle{spbasic}
\bibliography{fcrawford}


\end{document}